\documentclass{icrc2009}

\usepackage{graphicx}   
\usepackage[caption=false]{caption}    
\usepackage[font=footnotesize]{subfig} 
\usepackage{fixltx2e}
\usepackage{url}

\newcommand{\shorttitle}[1]%
{\markboth{Proceedings of the 31\MakeLowercase{$^{st}$} ICRC, {\L}\'{o}d\'{z} 2009}{#1} }
\newcommand{\etal}{\MakeLowercase{\textit{et al. }}} 

\begin{document}
\title{Fermi LAT Discovery of Gamma-ray Pulsars in a Blind Search}

\author{\IEEEauthorblockN{P.~M.~Saz Parkinson, M.~Dormody, and M.~Ziegler, for the Fermi-LAT Collaboration}
                            \\
\IEEEauthorblockA{Santa Cruz Institute of Particle Physics, U. of California, 1156 High Street, Santa Cruz, CA 95064, USA}}

\shorttitle{Saz Parkinson \etal Fermi-LAT Blind Search for Pulsars}
\maketitle

\begin{abstract}
The Large Area Telescope (LAT) on the Fermi Gamma-ray Space Telescope (formerly GLAST), with its improved sensitivity relative to previous generation 
gamma-ray telescopes, is significantly increasing the number of known gamma-ray sources in the sky, including pulsars. In addition to searching for gamma-ray 
emission from known radio pulsars, it is now possible to successfully perform blind searches for pulsars on the gamma-ray data alone, with the goal of uncovering 
a new population of potentially radio-quiet (Geminga-like) pulsars. We describe our methods and some recent results from our searches, performed using the 
time-differencing technique, which have resulted in the discovery of a large number of new gamma-ray pulsars.
  \end{abstract}

 
\section{Introduction}
Prior to the launch of $Fermi$, only six pulsars were known to emit $>$ 100 MeV gamma rays. The high energy pulsations from these spinning neutron stars were all detected by 
folding the gamma-ray data with the known ephemeris from radio observations, or in the case of the radio-quiet Geminga, using the X-ray ephemeris. In addition to 
these six gamma-ray pulsars, the EGRET mission on the \emph{Compton Gamma Ray Observatory} (1991-2000) detected hundreds of sources, a majority of which remained 
unidentified~\cite{Thompson}. Many unidentified gamma-ray sources, in particular those in the Galactic plane, have long been suspected of being pulsars, but deep radio 
and X-ray searches for pulsations have often come up empty. One of the key scientific objectives of $Fermi$ is to resolve the gamma-ray sky and identify these sources. In the first 
months of the mission, $Fermi$ is meeting those expectations and uncovering an entire population of pulsars hitherto unknown at any wavelength.\\
 
\section{The Fermi Large Area Telescope (LAT)}

The \emph{Fermi} satellite, consisting of the Gamma-ray Burst Monitor (GBM) and the Large Area Telescope (LAT), was launched on 11 June 2008 into a 
low Earth circular orbit at an altitude of 565 km and an inclination of 25.6$^\circ$. The LAT~\cite{atwood} is a pair-production telescope with large effective 
area ($\sim$8000 cm$^2$) and field of view (2.4 sr), sensitive to gamma rays between 20 MeV and $>$ 300 GeV. Although its commissioning phase 
(30 June to 30 July 2008) was primarily intended for instrument checkout and calibration, several scientific results were obtained with 
these early data, including the first discovery of a radio-quiet gamma-ray pulsar in a blind search~\cite{cta1}. The LAT began nominal science operations on 11 August 2008, 
and has since been observing mostly in survey mode, scanning the entire sky every three hours. The overall sensitivity of the LAT is $\sim$25 times that of 
EGRET, while the angular resolution is also significantly improved (it ranges from $\sim$3--6$^{\circ}$ at 100 MeV to $\sim$0.1--0.2$^{\circ}$ at 10 GeV). The mission was 
designed with a five-year lifetime and a goal of at least ten years of operations. The scientific goals of 
the mission span a wide range of topics, including understanding the emission mechanisms in Active Galactic Nuclei (e.g.~\cite{hayashida,raino,reyes}), 
pulsars (e.g.~\cite{celik,gargano,razzano}), and supernova remnants (SNR) (e.g. \cite{funk,lemoine,tanaka}), the high energy emission from Gamma-ray Bursts 
(e.g.~\cite{mcenery,omodei,tajima}), and the nature of dark matter (e.g.~\cite{bloom,nuss,profumo}).

\section{Blind Searches for Gamma-Ray Pulsars}

Detecting gamma-ray pulsars is challenging for a variety of reasons. The low count rates require long integration times (up to several years) and result in
very sparse data sets. This, combined with the short spin periods of most pulsars ($<$1s), presents serious computational problems for 
standard fully coherent search techniques involving Fast Fourier Transforms (FFT). In addition, during such long time spans, pulsars, especially young ones, will often experience 
irregularities in their timing behavior, such as ``timing noise" or glitches (e.g. PSR B1706-–44~\cite{pablo}), further complicating the problem. 

The term ``blind" refers to the fact that the spin parameters of the potential pulsar (e.g. frequency, $f$, 
and frequency derivative, $\dot{f}$) are unknown\footnote{Unlike in radio searches, we need not be concerned with searching over a range of values of dispersion measure (DM).}. 
A relatively small number of locations in the sky are searched for pulsations. These include promising sources from multiwavelength observations, as well as the 
locations of LAT-detected gamma-ray sources. A list of the most significant LAT sources detected in the first three months ($>10\sigma$) was recently released to the 
public~\cite{bsl}. All those not associated with AGN were searched. In addition, we searched other less significant LAT sources, from an internal catalog, which we deemed interesting 
pulsar candidates. At least one pulsar was found from a LAT source not included in the bright source list~\cite{science}.

The arrival times of photons must be translated to the Solar System barycenter assuming a location, and for each source, a large 
parameter space in $f$ and $\dot{f}$ must be covered. Using the Crab to define the upper limit for young pulsars, we chose the maximum frequency of 64 Hz and a 
minimum ratio of $\dot{f}$/$f$=-1.25$\times10^{-11}$ s$^{-1}$ (the parameter space is explored in steps of $\dot{f}$/$f$). These parameters cover roughly 85\% of 
the $\sim$2000 pulsars contained in the ATNF database. 

Figure 1 shows a LAT gamma-ray image of the Cygnus region of the Galactic plane, and serves to illustrate the different ways in which the LAT is finding gamma-ray 
pulsars. The bright source at the bottom is the gamma-ray counterpart of the known radio pulsar PSR J2021+3651 (also known as the ``Dragonfly"). This was the first gamma-ray pulsar 
detected by the LAT~\cite{dragonfly}, and indeed the first new ``post-EGRET" gamma-ray pulsar, although the original discovery was made slightly earlier by AGILE~\cite{halpern}. 
As in the case of the six EGRET pulsars, the gamma-ray pulsations were detected by folding the gamma-ray photons using the known radio ephemeris 
of the pulsar. Above the ``Dragonfly" pulsar lies LAT PSR J2021+4044 in the Gamma Cygni SNR. This source had long been suspected of being a 
pulsar and deep Chandra observations led to a list of possible X-ray counterparts~\cite{weisskopf}, which we used to search for, and find, the gamma-ray pulsar. Finally, in 
the top right corner of Figure 1, LAT PSR J2032+4127 is a pulsar coincident with the Cygnus OB2 association. The pulsations in this case were discovered in a blind search using the 
location of the LAT source. These three pulsars, like the majority of the gamma-ray pulsars discovered in our blind searches so far, are coincident with formerly unidentified 
EGRET gamma-ray sources: 3EG J2021+3716, 3EG J2020+4017, and 3EG J2033+4118 respectively. Furthermore, LAT PSR J2032+4127 is also coincident with the first unidentified TeV source, 
discovered originally by the HEGRA experiment~\cite{hegra}.

 \begin{figure}[!t]
  \centering
  \includegraphics[width=2.5in]{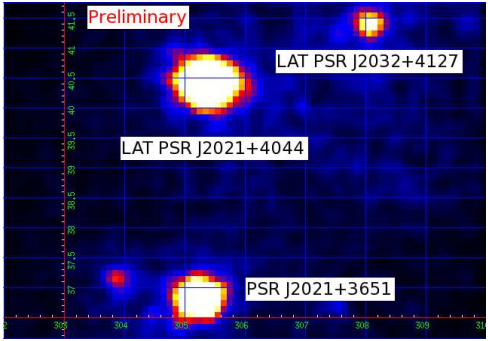}
  \caption{LAT gamma-ray image, in celestial coordinates, of the Cygnus region of the Galactic plane, showing three bright sources, coincident with formerly unidentified EGRET sources 3EG J2021+3716, 3EG J2020+4017, and 3EG J2033+4118. All three are new gamma-ray pulsars: the known radio pulsar PSR J2021+3651~\cite{dragonfly} and 
two gamma-ray pulsars discovered in the blind search, LAT PSR J2021+4044 in the Gamma Cygni SNR and LAT PSR J2032+4127 in the Cygnus OB2 association~\cite{science}.}
  \label{simp_fig1}
 \end{figure}

\subsection{The time-differencing technique}

\begin{figure}[!t]
  \centering
  \includegraphics[width=2.5in,angle=-90]{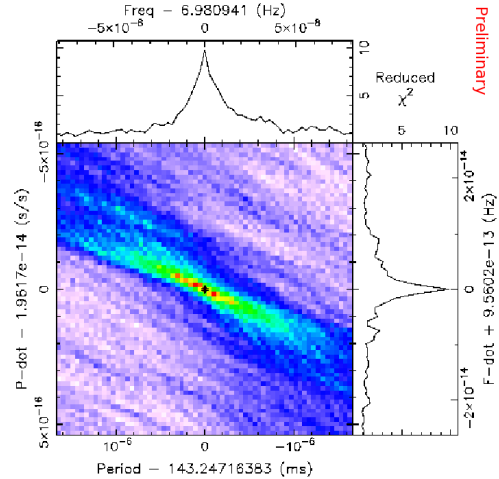}
  \caption{LAT PSR J2032+4127 contours of detection significance over a range of period and period derivative, as obtained by running the ``prepfold" command on PRESTO~\cite{presto}. A high value of reduced $\chi^2$ is shown in red and a lower value in white or purple.~\cite{science}.}
  \label{simp_fig2}
 \end{figure}

\begin{figure}[!t]
  \centering
  \includegraphics[width=2.5in]{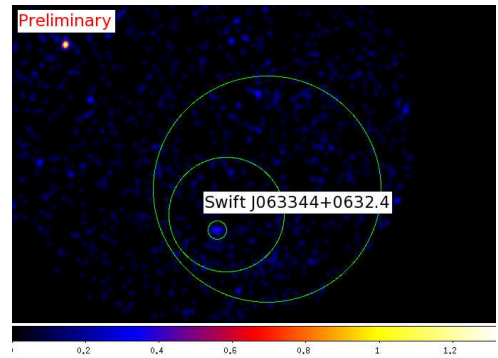}
  \caption{A $\sim$5 ks $Swift$ observation showing an X-ray image of the region of the sky, in celestial coordinates, around LAT PSR J0633+0632. The large circle represents the LAT 95\% error circle for the source after three months, while the smaller circle is the LAT 95\% error circle after 6 months. The crosshair marks the location of the assumed X-ray counterpart of the pulsar, which we have named $Swift$ J063344+0632.4~\cite{science}.}
  \label{simp_fig3}
 \end{figure}

 \begin{figure*}[!t]
   \centerline{\subfloat[Phase plot obtained using the LAT source location.]{\includegraphics[width=2.5in]{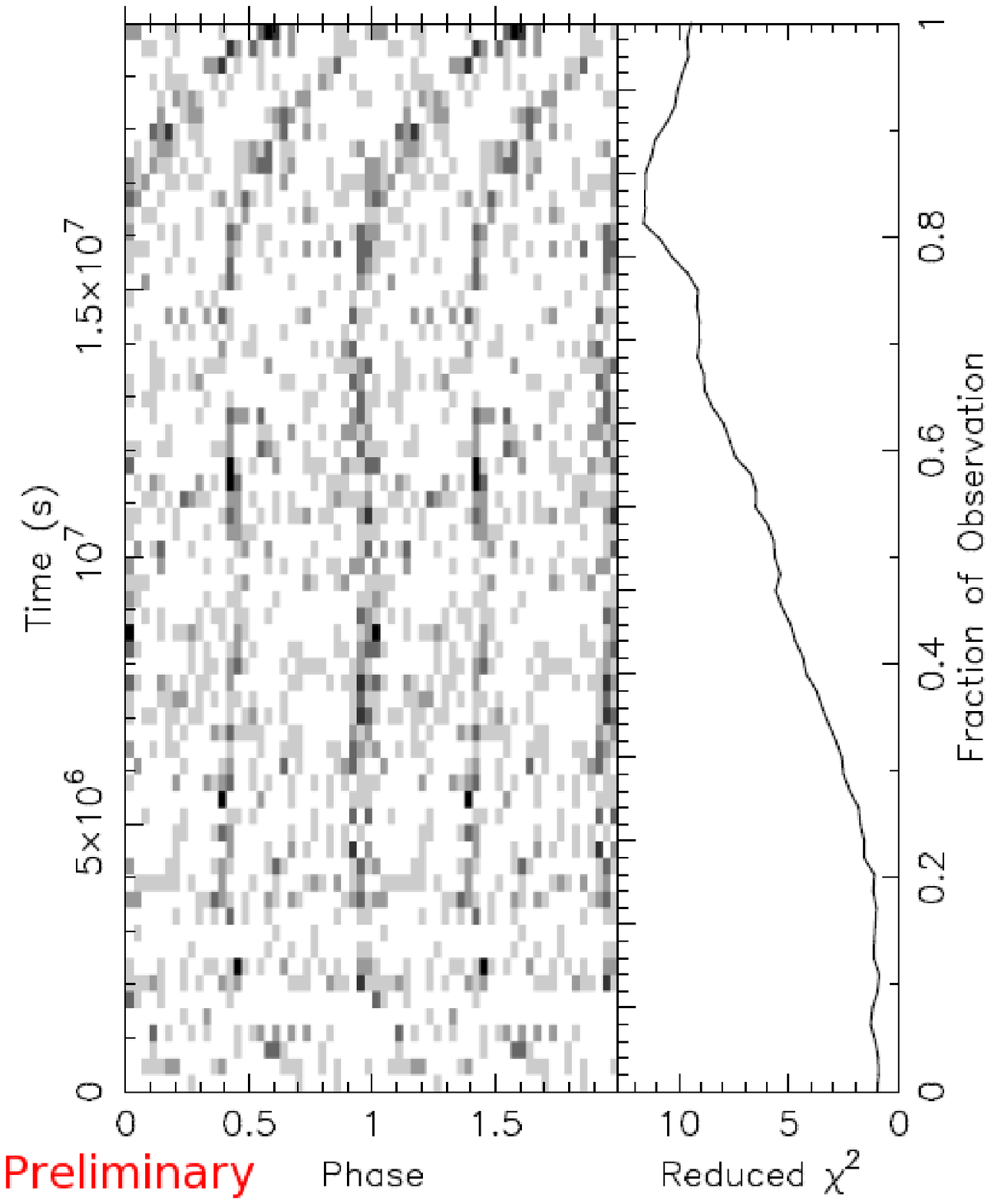} \label{sub_fig1}}
              \hfil
              \subfloat[Phase plot obtained using the $Swift$ source location.]{\includegraphics[width=2.5in]{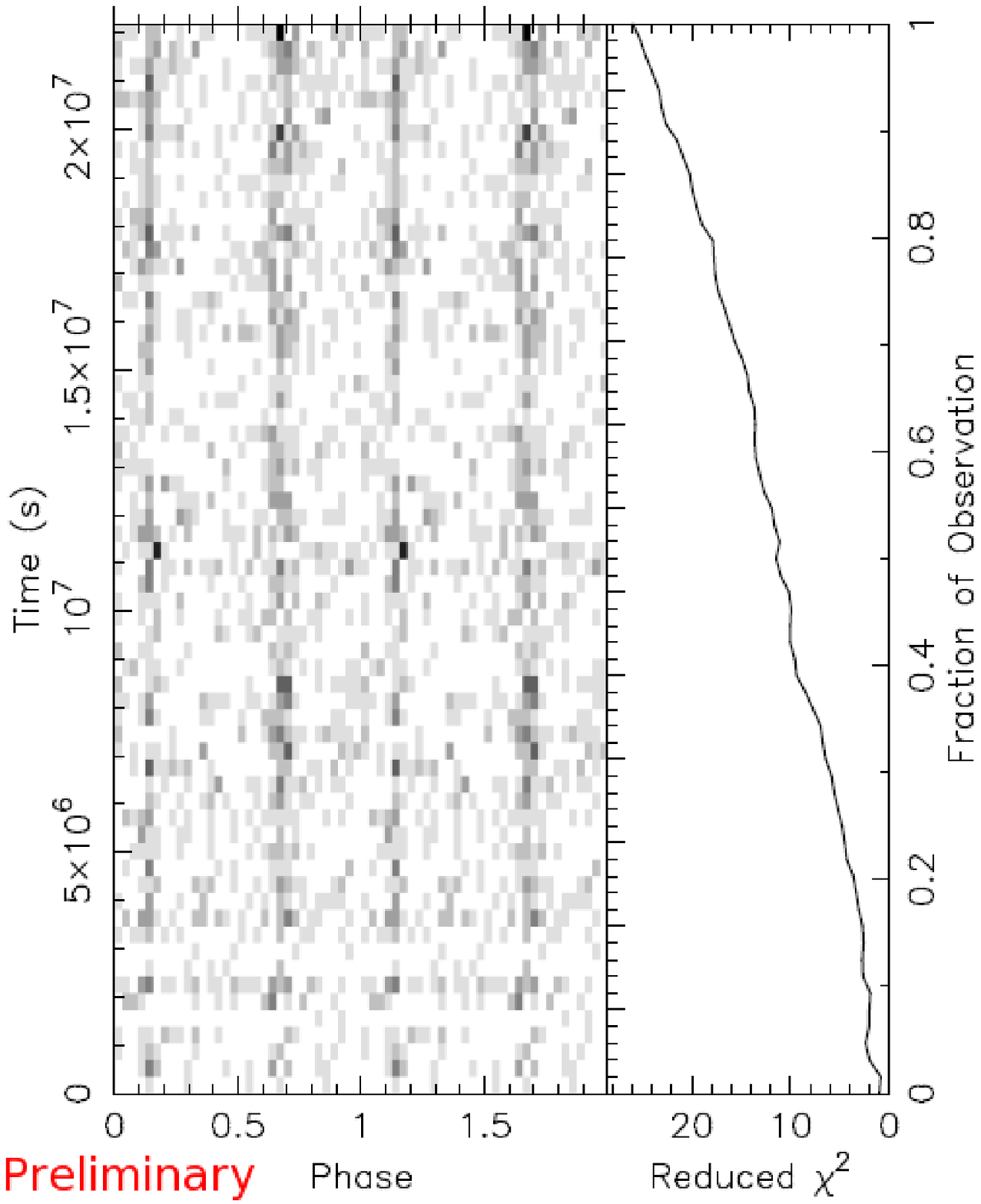} \label{sub_fig2}}
             }
   \caption{LAT PSR J0633+0632 phase plots and evolution of the reduced $\chi^2$ with time, using PRESTO~\cite{presto}. The plots illustrate the strength of the signal over the course of the entire observation, and the crucial importance of a precise location in determining a good timing solution.}
   \label{double_fig1}
 \end{figure*}

As described in the previous section, the low fluxes typical of gamma-ray sources result in very long, sparse data sets. The application of 
traditional FFT techniques on such data quickly become unfeasible, for several reasons. First, the frequency resolution from such long viewing periods results in FFTs with 
billions (or tens of billions) of frequency bins. More importantly, the relatively large frequency derivatives of pulsars means that these FFTs would have 
to be computed a large number of times in order to span a realistic parameter space, given the small step size in $\dot{f}$ required to keep the signal 
power within a single FFT bin. To make matters worse, the long viewing periods increase the chances that the pulsar will experience a glitch during this time. 
A new technique, known as ``time-differencing"~\cite{atwood06,ziegler08}, has been developed, which greatly improves the 
prospects of carrying out these searches. Based on the premise that a periodic signal present in a time series should also show up when analyzing the 
differences of these times, the method involves taking time differences only up to a short time window (on the order of days, rather than months or 
years). In doing this, the required number of FFT bins, $N$, is greatly reduced ($N = 2 \times f \times T_{w}$, where $f$ represents 
the maximum frequency searched and $T_{w}$ is the maximum time-difference window). The reduced frequency resolution also results in a larger step 
size required for $\dot{f}$, greatly reducing the number of $\dot{f}$ trials needed. The overall result is that the computational 
and memory costs become a fraction of the standard FFT methods, with only a small reduction in sensitivity. 
This makes it possible to carry out a large number of blind searches over a realistic parameter space on standard desktop computers with only a 
few GB of memory. That said, the blind search efforts have been significantly enhanced by access to the UCSC Astrophysics Supercomputer Pleiades, 
with 207 Dell PowerEdge 1950 compute nodes (828 processing cores). Despite the efficiency of the time-differencing technique, it is still 
desirable to limit the number of source locations in the sky that are searched, given both the finite computational resources available, and also to avoid 
incurring large penalties in the significance of our detections, due to the increasing number of trials used in finding a particular signal.

After a candidate pulsar is identified in our search, we use standard pulsar tools, such as PRESTO and TEMPO to refine the ephemeris. Figure 2 shows 
the results obtained by running the {\tt prepfold} command on PRESTO~\cite{presto}, showing the significance of the detection.
Often, the relatively large uncertainty in the LAT location makes it difficult to obtain a good timing solution. We then use several techniques to try to refine the
location of the source, the most fruitful of which involves multiwavelength observations. We have been fortunate to obtain $Swift$ follow-up
observations of several LAT pulsars, leading to the identification of potential X-ray counterparts. Figure 3 shows the result of a short $Swift$ observation of one of 
our newly-discovered pulsars, LAT PSR J0633+0632. The largest circle represents the 95\% error circle derived by the LAT after three months of data. The 
medium circle shows the 95\% error circle after six months of data. Finally, the crosshair pinpoints the X-ray source we believe to be the counterpart to the pulsar, a
source we have named $Swift$ J063344+0632.4. Figure 4 illustrates the crucial importance of having a precise location. The timing solution used in the left 
panel was the best we were able to obtain (using only a frequency and frequency derivative) for the LAT 6-month location. The panel on the right represents the best timing 
solution when using the location of the $Swift$ X-ray source to barycenter the gamma-ray photons.

\section{A new population of gamma-ray pulsars}

The first major $Fermi$ discovery was the detection of pulsations from CTA 1 in a blind search~\cite{cta1}. This young, nearby, shell-type SNR was 
discovered in radio in the 1960s and X-ray observations show a well-localized central point source, RXJ0007.0+7303, embedded in a pulsar wind nebula (PWN). 
High energy ($>$ 100 MeV) emission was detected by EGRET from 3EG J0010+7309, coincident with this source, but the low number of photons and poor 
localization made the search for pulsations very challenging. The new pulsar, with a period of 315.8637050 ms and a period derivative of 
3.615$\times$10$^{-13}$ s s$^{-1}$ is a typical young, energetic pulsar, with a derived characteristic age of $\sim$14,000 years 
(consistent with the estimated age of the SNR) and a spin-down power of 4.5$\times$10$^{35}$ erg s$^{-1}$. For more details, see \cite{cta1}.
Two other pulsars, LAT PSR J1809-2332 (``Taz") and LAT PSR J1836+5925 (the ``next" Geminga) were discovered in our blind searches during the commissioning phase of the instrument. 
All three of these pulsars were associated with former EGRET unidentified sources and in all cases it was long-suspected that a pulsar was responsible for the 
gamma-ray emission. Furthermore, the existence of X-ray sources thought to be the counterparts of such pulsars facilitated the detection of their gamma-ray 
pulsations. LAT PSR J1836+5925 has been named as
one of the top 5 most interesting Galactic sources detected by $Fermi$ so far~\footnote{http://www.nasa.gov/mission\_pages/GLAST/news/gammaray\_best.html}.
Our blind searches have continued to yield new and exciting detections of gamma-ray pulsars, to the point where there are now over a dozen new pulsars
discovered through their gamma-ray pulsations~\cite{science}. It is too early to say how many of these pulsars will turn out to be radio-quiet. Follow-up observations at the 
world's largest radio telescopes are in progress. The results of such searches will have important consequences in the population studies of pulsars and in our
understanding of the underlying pulsar emission mechanisms.

 \begin{figure}[!t]
  \centering
  \includegraphics[width=2.5in]{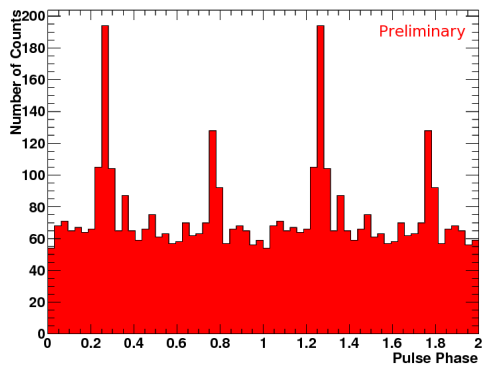}
  \caption{Folded light curve of LAT PSRJ2032+4127, with 32 phase bins per period, using 5 months of data with E$>$300 MeV, and R$<$0.8$^{\circ}$. Two rotations are shown for clarity~\cite{science}.}
  \label{simp_fig4}
 \end{figure}

\section{Conclusions}  
The $Fermi$ LAT has been scanning the gamma-ray sky with unprecedented sensitivity since its launch in June 2008. One of its key scientific missions is to 
resolve the gamma-ray sky by identifying gamma-ray sources that previous experiments left unidentified, and to discover new sources. The hugely improved sensitivity and angular resolution of the LAT, 
combined with a new powerful time-differencing technique, makes it possible, for the first time, to carry out sensitive searches for gamma-ray pulsations 
from all these sources. Even before the commissioning phase was complete, \emph{Fermi} already yielded important scientific discoveries in the field of pulsar astrophysics, 
including the detection of gamma-ray pulsations from the "Dragonfly" pulsar (PSR J2021+3651)~\cite{dragonfly}, the detection of the first gamma-ray glitch from 
PSR B1706--44~\cite{pablo}, and the discovery of a new radio-quiet gamma-ray pulsar in the SNR CTA 1~\cite{cta1}. With the discovery of a large number of additional gamma-ray pulsars~\cite{science}, \emph{Fermi} is fulfilling its pre-launch expectations by identifying a large number of formerly unidentified gamma-ray sources and at the same time uncovering a 
whole new population of pulsars never before accessible by any other means.

\section*{Acknowledgements} 
I am grateful for the support of the American Astronomical Society and the National Science Foundation in the form of an International Travel Grant, which enabled me to attend 
this conference. 

The $Fermi$ LAT Collaboration acknowledges support from a number of agencies and institutes for both development and the operation of 
the LAT as well as scientific data analysis. These include NASA and DOE in the United States, CEA/Irfu and IN2P3/CNRS in France, ASI 
and INFN in Italy, MEXT, KEK, and JAXA in Japan, and the K.~A.~Wallenberg Foundation, the Swedish Research Council and the National 
Space Board in Sweden. Additional support from INAF in Italy for science analysis during the operations phase is also gratefully 
acknowledged.


\begin{thebibliography}{99}
\bibitem{Thompson} Thompson, D.~J., Reports on Progress in Physics, 71, 116901 (2008)
\bibitem{atwood} Atwood, W. B., et al., \emph{The Large Area Telescope on the Fermi Gamma-ray Space Telescope}, ApJ, in press, arxiv:0902.1089 (2009)
\bibitem{cta1} Abdo, A.~A., et al., Science, 322, 1218 (2008) 
\bibitem{hayashida} Hayashida, M., these proceedings (2009)
\bibitem{raino} Rain\`{o}, S., these proceedings (2009)
\bibitem{reyes} Reyes, L., these proceedings (2009)
\bibitem{celik} Celik, O., these proceedings (2009)
\bibitem{gargano} Gargano, F., these proceedings (2009)
\bibitem{razzano} Razzano, M., these proceedings (2009)
\bibitem{funk} Funk, S., these proceedings (2009)
\bibitem{lemoine} Lemoine, M., et al, these proceedings (2009)
\bibitem{tanaka} Tanaka, T., these proceedings (2009)
\bibitem{mcenery} McEnery, J., these proceedings (2009)
\bibitem{omodei} Omodei, N., these proceedings (2009)
\bibitem{tajima} Tajima, H., these proceedings (2009)
\bibitem{bloom} Bloom, E., et al., these proceedings (2009)
\bibitem{nuss} Nuss, E., et al., these proceedings (2009)
\bibitem{profumo} Profumo, S., et al., these proceedings (2009)
\bibitem{pablo} Saz Parkinson, P.~M.\, American Institute of Physics Conference Series, 1112, 79 (2009)
\bibitem{bsl} Abdo, A.~A., et al., \emph{Fermi Large Area Telescope Bright Gamma-ray Source List}, ApJS, submitted, arXiv:0902.1340 (2009)
\bibitem{science} Abdo, A.~A., et al., \emph{Sixteen Gamma-Ray Pulsars Discovered in Blind Frequency Searches Using the Fermi LAT}, Science, submitted (2009)
\bibitem{dragonfly} Abdo, A.~A., et al., \emph{Pulsed gamma-rays from PSR J2021+3651 with the Fermi Large Area Telescope}, ApJ, submitted (2009)
\bibitem{halpern} Halpern, J.~P., et al., ApJL, 688, L33 (2008)
\bibitem{weisskopf} Weisskopf, M.~C., et al., ApJ, 652, 387 (2006)
\bibitem{hegra} Aharonian, F., et al, A\&A, 393, L37 (2002) 
\bibitem{atwood06} Atwood, W.~B., Ziegler, M., Johnson, R.~P., \& Baughman, B.~M., ApJL, 652, L49 (2006) 
\bibitem{ziegler08} Ziegler, M., Baughman, B.~M., Johnson, R.~P., \& Atwood, W.~B., ApJ, 680, 620 (2008) 
\bibitem{presto} Ransom, S.~M., PhD Thesis, Harvard University (2001)








  \end{thebibliography}
\end{document}